\documentclass[aps,twocolumn,prl,groupedaddress,showpacs]{revtex4-1}

\usepackage{graphicx}

\begin{document}


\title{Dark states of single NV centers in diamond unraveled by single shot NMR}


\author{G.~Waldherr}
\email{g.waldherr@physik.uni-stuttgart.de}
\author{J.~Beck}
\author{M.~Steiner}
\author{P.~Neumann}
\author{A.~Gali}
\altaffiliation{Research Institute for Solid State Physics and Optics, Hungarian Academy of Sciences, 1525 Budapest, Hungary}
\author{Th. Frauenheim}
\altaffiliation{Bremen Center for Computational Materials Science, University of Bremen, 28359 Bremen, Germany}
\author{F.~Jelezko}
\author{J.~Wrachtrup}
\affiliation{3. Physikalisches Institut and Research Center SCOPE, University of Stuttgart, 70569 Stuttgart, Germany}


\date{\today}

\begin{abstract}

The nitrogen-vacancy (NV) center in diamond is supposed to be a building block for quantum computing and nanometer scale metrology at ambient conditions.
Therefore, precise knowledge of its quantum states is crucial.
Here, we experimentally show that under usual operating conditions the NV exists in an equilibrium of two charge states (70\% in the expected negative (NV$^-$) and 30\% in the neutral one (NV$^0$)).
Projective quantum non-demolition measurement of the nitrogen nuclear spin enables the detection even of the additional, optically inactive state.
The nuclear spin can be coherently driven also in NV$^0$ ($T_1 \approx 90$~ms and $T_2 \approx 6$~$\mu$s).
\end{abstract}

\pacs{76.30.Mi, 03.67.-a, 78.55.Cr}

\maketitle

\begin{figure}
\includegraphics[width=8.5cm]{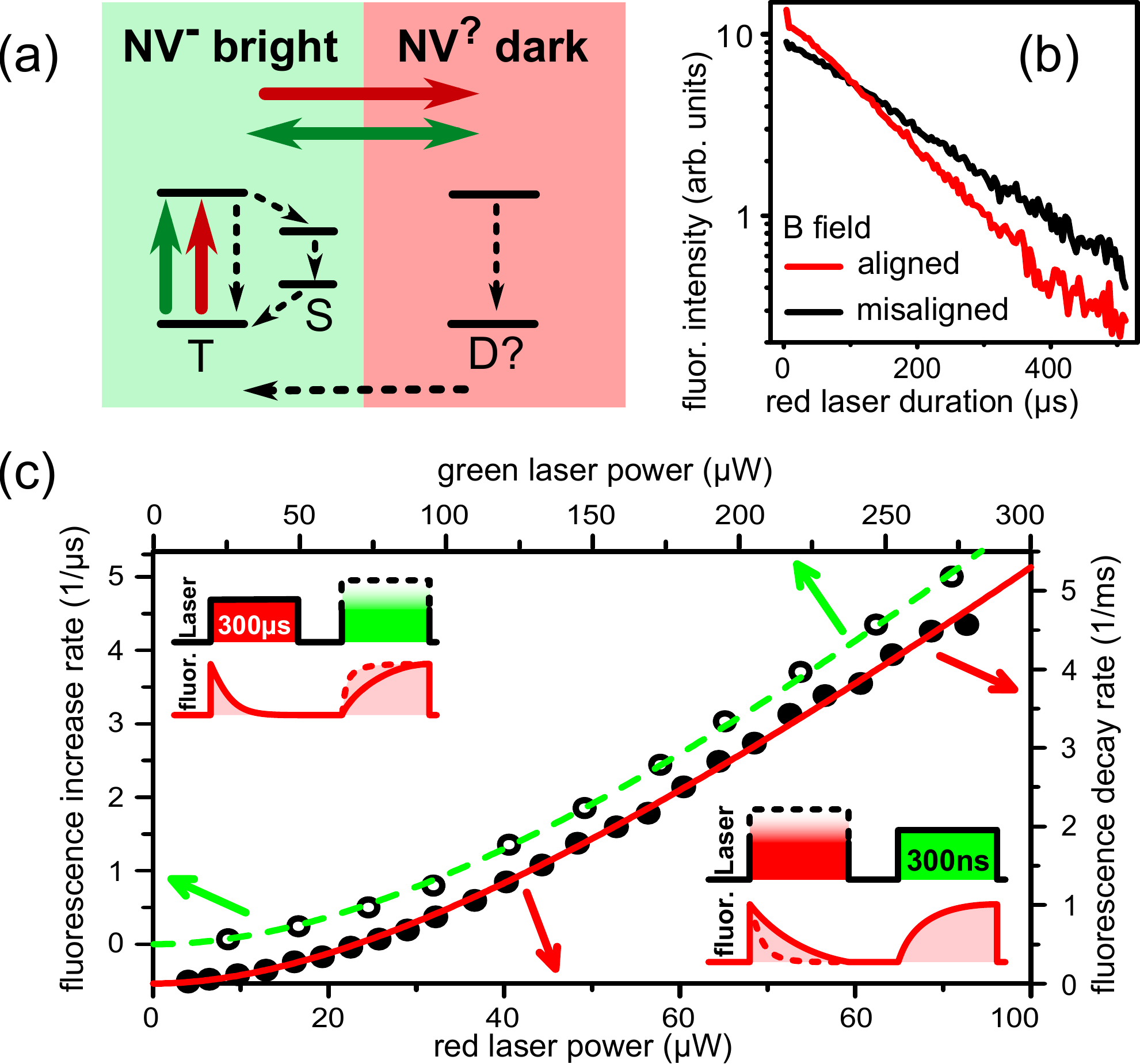}
\caption[optical transitions between bright and dark state]{\label{fig:2}(a) Energy level scheme of bright and dark state. The bright state comprises of the triple (T) and singlet (S) levels of NV$^{-}$ \cite{Mans_Arxiv_2010} whereas the dark state is at least a doublet (D) supposedly in another charge state. The arrows symbolize transition rates, where green and red are due to the respective lasers and the black dashed arrows are decay processes. (b) The fluorescence signal of the bright state for increasing red laser duration shows an exponential decay depending on the magnetic field orientation (aligned$\rightarrow\tau=120$~$\mu$s, misaligned $\rightarrow\tau=184$~$\mu$s) 
(c) Fluorescence decrease and increase rates depending on corresponding red or green laser power for pumping into or out of the dark state respectively. Insets show respective pulse sequences and corresponding fluorescence intensities. For fits see text.}
\end{figure}

Electron and nuclear spins of the NV color center in diamond are promising candidates for solid state quantum information processing \cite{Gaeb_NatPhys_2006,Dutt_Science_2007,Neum_Science_2008,Capp_PRL_2009,Jian_Science_2009,Fuch_Science_2009,Neum_NatPhys_2010,Toyl_NanoLett_2010} and field sensing at nanometer scale under ambient conditions \cite{Bala_Nature_2008,Maze_Nature_2008}.
It consists of a substitutional nitrogen atom next to a lattice vacancy and if negatively charged exhibits an electron spin triplet ground state with very favorable coherence properties \cite{Grub_Science_1997,Bala_NatMater_2009,Lang_Science_2010}.
Furthermore, electron and nuclear spins can be efficiently initialized and readout optically even at room temperature.
\\
\indent By optically pumping with laser light in the green spectral range the NV$^-$ center is initialized into the $m_S=0$ spin sublevel of its triplet ($S=1$) ground state.
It is this state that exhibits a higher fluorescence rate than the $m_S\pm1$ states which is vital for optical spin state detection.
At the same time the electronic spin is polarized to a degree far beyond thermal equilibrium which is a necessity for quantum computing and magnetometry.
All these features belong to the negative charge state of the NV center which is referred to as ``bright state'' in the rest of this letter (fig. \ref{fig:2}(a)).
However, the NV center can also be present in different charge states.
Depending on the Fermi level other charge states can be more stable such as the neutral one (NV$^0$) which is associated with fluorescence with a zero phonon line at 575~nm and a phonon sideband up to 700~nm \cite{Rond_PRB_2010}.
It was shown that laser irradiation can induce interconversions between NV$^-$ and NV$^0$ \cite{Mans_Diam_2005,Gaeb_APB_2006}.
Furthermore, recently a long living (150~s) ``dark state'' was observed \cite{Han_NanoLetters_2010}, which was proposed to be a metastable singlet state of NV$^-$.
Owing to applications as quantum bit the precise characterization and control of the NV center initial state is crucial.
\\
\indent In this letter we show that quantum non-demolition (QND) measurement of single nitrogen nuclear spins at room temperature can be used to probe different charge states of the NV center.
We provide a quantitative analysis of populations in different NV center charge states.
It will turn out that next to the negative charge state (bright state) the NV center resides a substantial amount of time in the dark state.
Although it is optically inactive it can be detected by the characteristic hyperfine coupling to the nuclear spin.
\\
\indent The experimental setup consists of a homebuilt room temperature microscope adapted for single spin magnetic resonance.
It allows projective read out of a single nuclear spin in a ``single shot'' and initialization of the spin in a desired pure state as recently discovered \cite{Neum_Science_2010}.
Thus, now we are able to correlate the absolute numbers of attempts to flip a spin and real spin flips.
The experiments have been performed on several single, natural abundant NV centers in bulk and nano-diamonds (average size: 50~nm).
\begin{figure}
\includegraphics[width=8.5cm]{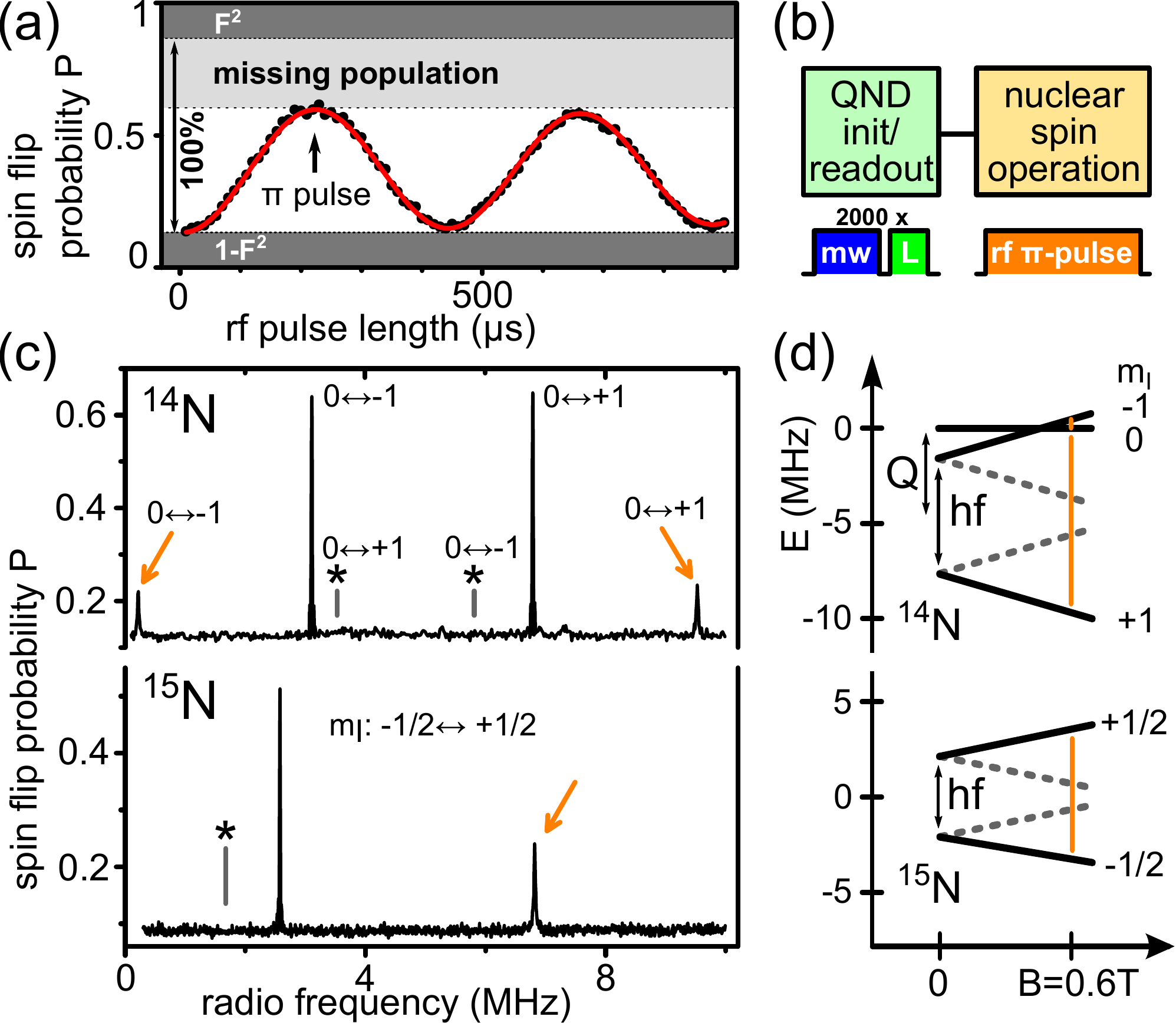}%
\caption[NMR spectra]{\label{fig:1}(a) Nitrogen nuclear spin Rabi oscillation (NV$^-$, $m_S=0$). The fidelity $F$ reduces the achievable contrast ($\rightarrow100\%$ population) by the dark gray regions. The even smaller Rabi amplitude is because of some population that is not in the NV$^-$ $m_S=0$ state (light gray region). 
(b) QND initialization and readout of nuclear spin state: microwaves (mw) manipulate the electron spin, green Laser (L, 532nm) reads it out and polarizes it into $m_S=0$ \cite{Neum_Science_2010}. A resonant rf $\pi$-pulse flips the nuclear spin.
(c) NMR spectra showing transition frequencies of the $^{14}$N and $^{15}$N nuclear spin for different NV center states. The corresponding quantum numbers $m_I$ for each transition are given. Dark state NMR transitions are marked with orange arrows; other lines belong to the bright state $m_S=0$. Gray lines with stars mark expected transitions in the dark state (see text). (d) Energy level scheme of NV center nuclear spin in dark state. NMR transitions among spin states are marked with orange lines (arrows in (c)). Hyperfine interaction $hf$ and quadrupole splitting $Q$ are marked. The gray dashed lines mark the energy levels for a reversed hyperfine field (stars in (c)).}
\end{figure}
Figure \ref{fig:1}(a) shows a typical Rabi oscillation of the $^{14}$N nuclear spin ($I=1$).
The vertical axis shows the probability for a spin flip induced by radio frequency (rf).
The limited readout and initialization fidelity $F$ reduces the contrast of measurement symmetrically (i.e. full oscillations start at $1-F^2$ and go up to $F^2$).
Remarkably, the visible oscillations are not symmetric with respect to the 50/50 spin flip probability line indicating that in 30$\pm$3\% (for all investigated NVs) of all cases the radio frequency does not flip the nuclear spin even for well adjusted $\pi$-pulses and high Rabi frequencies.
This is because the nuclear spin is out of resonance in the latter cases.
We attribute this effect to the NV center being in an unknown state (which might also be another charge state) with a different hyperfine field.
\\
\indent To identify this state, NMR spectra are taken for single NV centers with nitrogen isotopes $^{14}$N ($I=1$) and $^{15}$N ($I=1/2$) by performing the following sequence: (\textit{1}) QND nuclear spin state initialization into $m_I=0$ ($^{14}$N) or $m_I=+1/2$ ($^{15}$N), (\textit{2}) nuclear spin $\pi$-pulse and (\textit{3}) QND nuclear spin readout (see fig. \ref{fig:1}(b)).
The applied magnet field was $B=0.6$T.
This procedure is repeated for a range of NMR frequencies and results in the spectra shown in figure \ref{fig:1}(c).
They show strong lines corresponding to nuclear spin flips of the NV$^-$ defect in its $m_S=0$ state (two lines for $^{14}$N and a single line for $^{15}$N).
Additionally, weaker lines are observed (marked by arrows) that belong to the unknown state.
From their spectral position and their dependence on magnetic field the nuclear spin energy level diagram in the unknown state can be deduced (fig.\ref{fig:1}(d)).
This reveals hyperfine ($hf$) coupling to an electronic magnetic moment $M$ which does not match the hyperfine coupling within the $S=1$ manifold of the NV$^-$ ground state.
Surprisingly, we see hyperfine coupling to only one projection of $M$ ($m_M=\mu$).
This might be due a polarized magnetic moment, however, we do not know the polarization mechanism.
At fields lower than $0.4$T also the other lines corresponding to $m_M=-\mu$ start to appear, which indicates a $M=1/2$ system.
In addition, we have observed that the lifetime of this unknown state is $\gg 1$~s.
Hence, we conclude that this state is a ground or long living metastable state.
\\
\indent The hyperfine coupling constants $a$ in the unknown state are deduced according to $hf=a \cdot m_M \cdot \Delta m_I$, where $hf$ is the hyperfine induced frequency splitting between two nuclear spin projections ($\pm\frac{1}{2}$ for $^{15}$N and $\pm1$ for $^{14}$N, cf. fig.\ref{fig:1}(d)) and $m_M$ is the projection of $M$.
For $^{15}$N we find $|a_{15\mathrm{N}} \cdot m_M|=4.242$~MHz and for $^{14}$N we find $|a_{14\mathrm{N}} \cdot m_M|=3.03$~MHz (in the case of $M=S=\frac{1}{2}$ this yields $|a_{15\mathrm{N}}|=8.484$~MHz and $|a_{14\mathrm{N}}|=6.06$~MHz).
In addition to hyperfine coupling, $^{14}$N exhibits a quadrupole splitting between nuclear spin projection $m_I=0$ and $m_I=\pm1$ of $Q=-4.654$~MHz (fig.\ref{fig:1}(d)).
\\
\indent The values above are on the same order as for the NV$^-$ center in its triplet ground state. 
It is expected that hyperfine coupling to any electron spin in the $\mathrm{e}_{x,y}$ orbitals of the NV center would be in this range (e.g. $S=1$ of NV$^{-}$, $S=1/2$ of NV$^{2-}$ or $S=1/2$ of NV$^0$ \cite{Felt_PRB_2009,Felt_PRB_2008}).
Hence, the unknown state might be the ground state of another charge state of the NV center exhibiting an electron spin.
Excited states of the aforementioned charge states would lead to spin population of the $\mathrm{a}_{\mathrm{N}}$ orbital, showing much larger hyperfine couplings \cite{Felt_PRB_2008}.
\\
\indent The interconversion between the different states of the NV center observed above can be controlled by intensity and wavelength of the applied laser radiation.
As can be seen green laser light (532~nm) establishes an equilibrium between the NV$^-$ $m_S=0$ state and the unknown state \ref{fig:1}(c).
Now we show that the unknown state is in fact the dark state observed in \cite{Han_NanoLetters_2010}.
It can be selectively populated by red laser illumination.
\begin{figure}
\includegraphics[width=8.5cm]{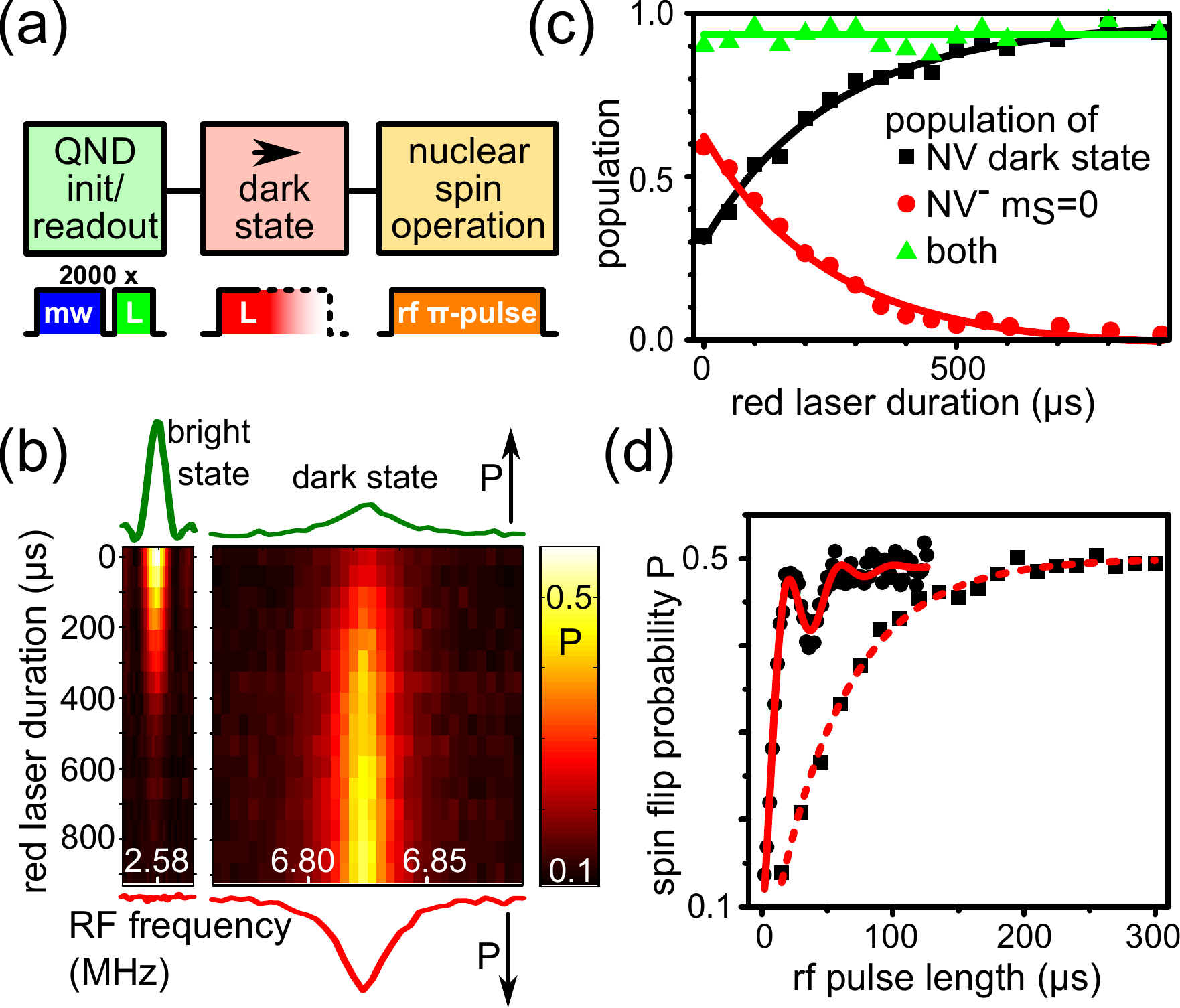}%
\caption[identification of dark state]{\label{fig:3}(a) NMR pulse sequence to investigate transition into the dark state. After readout and initialization a red laser pulse (637~nm) of variable length transfers a certain amount of population into the dark state. A subsequent nuclear spin $\pi$-pulse monitors the NMR line intensities. (b) 2D plot of NMR amplitude (color coded) over rf frequency (horiz.) and red laser pulse length (vert.). Spectra for minimal and maximal laser pulse lengths are shown above and below the 2D plot. (c) Populations in the dark and bright state deduced from (b). (d) Nuclear spin Rabi oscillation in the dark state for rf fields of two different strengths (18dB difference), showing homogeneous dephasing.} 
\end{figure}
Therefore, the NMR transitions of the $^{15}$N nuclear spin (see fig. \ref{fig:1}(c)) are monitored for an increasing length of the red laser (637~nm) pulses (i.e. for an increasing fraction of population in the dark state).
The corresponding measurement sequence is displayed in figure \ref{fig:3}(a) and the results in figure \ref{fig:3}(b).
It reveals that the NMR line corresponding to the bright state vanishes completely for increasing laser pulse length whereas the NMR line corresponding to the dark state starts with a finite amplitude and rises to a maximum.
From the example NMR line plots it is also visible that, whereas the bright state line shows the typical Fourier transformed lineshape due to the rectangular rf $\pi$-pulse the dark state line is a much broader Lorentzian (full width at half maximum $\approx 33$~kHz due to homogeneous line broadening (cf. fig.\ref{fig:3}(d)) indicating $T_2=T_2^*\approx6\mu s$).
In addition the nuclear spin lifetime $T_1$ is shorter in the dark than in the bright state ($90\mathrm{ms}<800\mathrm{ms}$, measured by monitoring the nuclear spin polarization for different periods in the dark state).
Coherent driving of the dark state NMR transition reveals fast decaying Rabi oscillations (fig.\ref{fig:3}(d)) towards 50\% of the initial amplitude irrespective of the Rabi frequency.
Thus, the amplitude in the dark state NMR lines in fig. \ref{fig:1}(c) represent only half of the population in the dark state.
From these results we can deduce the population in the NV center states for each laser pulse length (see figure \ref{fig:3}(c)).
Apparently, at least 95\% of the population is distributed among the bright state's $m_S=0$ sublevel and the dark state.
The degree of spin polarization within the NV$^{-}$groundstate is $\geq92$\%, which is higher than recent findings \cite{Neum_NatPhys_2010, Fuch_NatPhys_2010, Robl_arxiv_2010, Mans_Arxiv_2010}.
Furthermore, the red laser pulse seems to transfer nearly 100\% of the population into one $m_M$ level of the dark state.
\\
Further insight in the nature of the dark state can be deduced from purely optical experiments.
In the following, we investigate the kinetics of the optically induced transitions between dark and bright state (fig.\ref{fig:2}(a)).
The quadratic dependence of the transition rates on the laser power in both directions indicates a two-photon nature of the processes (fig.\ref{fig:2}(c)).
However, for stronger laser powers the dependence becomes linear which corresponds to a transition via a real excited state level that becomes saturated at higher powers.
Further evidence for an excited state assisted transition to the dark state can be found by changing the steady state population of the excited state while leaving the laser intensity constant.
This can be achieved by application of a strongly misaligned magnetic field, shelving the population from the excited state triplet into the metastable singlet state \cite{Epst_NatPhys_2005}.
As expected, the transfer rate to the dark state is reduced when the magnetic field is misaligned such that the fluorescence of the NV is minimized (fig.\ref{fig:2}(b)).
Another hint that we substantially populate the excited state triplet of NV$^-$ is its electron spin polarization into $m_S=0$ by optical excitation with the red laser.
This usually works by intersystem crossing from the excited state triplet into the singlet states (fig.\ref{fig:2}(a)).
The excited state of the NV$^-$ defect is located $\approx 0.6$~eV below the conduction band of bulk diamond \cite{Stee_Diam_2000}.
Thus, the excited state absorption is likely to ionize the defect leading to NV$^0$.
We assume that the excited electron becomes trapped by a donor (e.g. nitrogen impurity) \cite{Robl_PRL_2010}, which is not photo-ionized by red light.
The power dependence of the re-pumping rate by ionization of donors with green light indicates a similar two-photon process as explained above.
The ratio between rates to and from the dark state under green excitation should be the inverse population ratio (i.e. $\approx0.4$).
\\
\indent The magnetic moment $M$ of the dark state originates from either an electron spin or an orbital angular momentum of the NV center.
The position (and intensity) of the dark state NMR lines were identical for different NV centers even in nano diamonds where strain is particularly high.
Thus the nature of the magnetic moment $M$ is very likely related to an electron spin as orbital angular momentum is very strain sensitive.
Furthermore, it is known that the dynamic Jahn-Teller effect (DJT) in the NV center can average out orbital angular momentum at room temperature as shown for the NV$^-$ excited state \cite{Fu_PRL_2009}.
In addition, it is supposed to cause EPR line broadening in the NV$^0$ $S=1/2$ ground state \cite{Felt_PRB_2008,Gali_PRB_20091} and can therefore be responsible for the observed fast dephasing of the nuclear spin in our experiments.
All observations presented in this letter are in agreement with the assignment of the dark state to the NV$^0$ center, though future experiments would be needed for a definite proof.
\\
\indent Summarizing, with laser light of 532~nm and 637~nm we were able to selectively address different charge states of the NV center in diamond.
QND measurement enables the investigation of these states.
All this sheds new light on previous experiments using single NV$^-$ centers.
Whenever rate models were used to understand the fluorescence behavior and its spin state dependence \cite{Mans_Arxiv_2010,Fuch_NatPhys_2010,Robl_arxiv_2010} the dark state can now be incorporated for a more accurate picture.
Our results also make 70\% an upper bound on the fluorescence quantum yield of the investigated NV centers upon usual green illumination (not taking into account intersystem crossing that would further decrease the quantum yield).
In addition, previous spin operations have been only 70\% efficient (i.e. only if the NV resides in the bright state).
Surprisingly, QND initialization and readout of single nuclear spins is working even under this condition showing that the nuclear spin state is preserved during ionization and deionization.
Thus, it might be possible to detect shifts of the dark/bright state equilibrium when changing the Fermi level (e.g. in nano diamonds \cite{Rond_PRB_2010} or close to surfaces \cite{Hauf_arxiv_2010}) or when changing the dopant concentration of the diamond lattice.
In addition it was proposed to use the dark state for high resolution imaging \cite{Han_NanoLetters_2010}.
If there is a way to bring 100\% of the population into the bright state, NV$^-$ spin operations will be 100\% efficient.
This is of prime importance for numerous emerging applications of the NV defect center (e.g. quantum information processing and field sensing at the nanometer scale) \cite{Gaeb_NatPhys_2006,Dutt_Science_2007,Neum_Science_2008,Capp_PRL_2009,Jian_Science_2009,Fuch_Science_2009,Neum_NatPhys_2010,Bala_Nature_2008,Maze_Nature_2008}.
An alternative strategy is using the nuclear spin in the dark state as a qubit.
It seems possible to initialize it to 100\% and single shot readout is working as well via the bright state.
To improve quantum state lifetimes the application of uniaxial stress might be successful in case of DJT \cite{Felt_PRB_2008}.
\\
\indent We thank Roman Kolesov for experimental assistance and discussion as well as Petr Siyushev and Gopalakrishnan Balasubramanian for discussion.
Support by the Deutsche Forschungsgesellschaft (SFB/TR21, FOR 1482), the European Commission (SOLID, DIAMANT), the Volkswagenstiftung and the Landesstiftung BW is gratefully acknowledged.
\bibliography{references}

\end{document}